\documentclass[10pt, twocolumn, secnumarabic, amssymb, nobibnotes, showpacs, aps, pra,superscriptaddress]{revtex4-1}

\usepackage{graphicx}
\usepackage{dcolumn}
\usepackage{amsmath}
\usepackage{natbib}

\begin{document}

\title{Information-theoretic Bell inequalities based on Tsallis entropy}

\author{Marek Wajs}   \email{marek.wajs@nus.edu.sg}   \affiliation{Centre for Quantum Technologies, National University of Singapore, 3 Science Drive 2, 117543 Singapore, Singapore}

\author{Pawe\l{} Kurzy\'nski}   \email{cqtpkk@nus.edu.sg}   \affiliation{Centre for Quantum Technologies, National University of Singapore, 3 Science Drive 2, 117543 Singapore, Singapore} \affiliation{Faculty of Physics, Adam Mickiewicz University, Umultowska 85, 61-614 Pozna\'n, Poland}

\author{Dagomir Kaszlikowski}   \email{ phykd@nus.edu.sg}   \affiliation{Centre for Quantum Technologies, National University of Singapore, 3 Science Drive 2, 117543 Singapore, Singapore}

\date{\today}

\begin{abstract}
We numerically  investigate entropic Bell inequalities for a pair of entangled qutrits using information-theoretic distances. We show that for this class of inequalities Tsallis entropy is more suitable than Shannon as it reveals non-classicality for a larger set of quantum states. Finally, we find that like probability based inequalities, entropic ones are maximally violated by the non-maximally entangled qutrit state.

\end{abstract}

\pacs{Valid PACS appear here}

\maketitle

\section{Introduction}
Bell inequalities test whether measurements on spatially separated systems admit local-realistic description. Lack of thereof can be used in quantum information processing tasks such as device independent quantum key distribution \cite{Acin2006}, private randomness amplification \cite{Pironio2010}, quantum computation \cite{Nielsen2000} and many others.

In general, Bell inequality gives a local-realistic bound $b$ on some function $f$ of data $\{a_1,a_2,\dots \}$ that was acquired during a measurement process, i.e., $f(a_1,a_2,\dots ) \leq b$. For different choices of $f$ one obtains different types of inequalities, like probability based inequalities \cite{Clauser1974}, correlation inequalities \cite{Clauser1969}, or entropic inequalities \cite{Braunstein1988}.    

An interesting approach was proposed by Braunstein and Caves \cite{Braunstein1988}. In this case function $f$ is based on Shannon entropy. Later Schumacher showed that these Bell inequalities have an intuitive geometrical interpretation \cite{Schumacher1991}. What is more, application of information-theoretic distance measure allows for a unification of different types of inequalities, since an information-theoretic distance can be formulated in many ways  \cite{Schumacher1991}-\cite{Kurzynski2014}.

In this work we follow the information-theoretic distance approach and derive new Bell inequalities utilizing Tsallis entropy. Another approach to Bell inequalities in which Tsallis entropies are used was recently proposed in \cite{Manko2014}. We apply and analyze them numerically on a two-qutrit state. The motivation behind our research is twofold. Firstly, although entropic based inequalities were never shown to be tight (in a sense that the lack of their violation does not imply local realistic description), their advantage is that they do not rely on a labeling of measurement outcomes and that their form does not depend on a dimension of the system. This property is important for high dimensional quantum systems. 

The second motivation stems from the fact that entropic inequalities are not as robust against experimental noise as the ones constructed from probabilities and correlation functions \cite{Cerf1997}. To improve the noise tolerance we propose to use Tsallis entropy instead of Shannon entropy.  The advantage of the former is that it can be optimized with respect to the probability distribution of investigated variables.

We derive entropic Bell inequalities following the ideas in \cite{Kurzynski2014}. These inequalities utilise Tsallis entropy which increases its robustness against environmental white noise.  We also find that the optimal quantum state for the violation of these inequalities is maximally entangled. This is in line with the probability based inequalities \cite{Kaszlikowski2002}, \cite{CGLMP2002} for which the optimal quantum state is not maximally entangled.

\section{Theoretical background}

\subsection{Information-theoretic metric}

An information-theoretic metric can be defined between a pair of measurements that can be jointly performed. Consider two measurements  $X_i$ and $X_j$ with corresponding outcomes $x_i$ and $x_j$ and a joint probability distribution $p(x_i,x_j)$. We define a real function $d(X_i,X_j)$ of $p(x_i,x_j)$ that obeys non-negativity, symmetry and triangle inequality 
\begin{itemize}
\item $d(X_i,X_j) \geq 0$, ~~~$d(X_i,X_j)=0$ iff $X_i=X_j$;
\item $d(X_i,X_j)=d(X_j,X_i)$;
\item $d(X_i,X_j)+d(X_j,X_k) \geq d(X_i,X_k)$.
\end{itemize}

Derivation of distance-based Bell inequalities relies on one crucial assumption \cite{Kurzynski2014}: $d(A,B)$ exists even if $A$ and $B$ cannot be jointly measured. This assumption is valid in any local realistic theory for which joint probability distributions exists \cite{Fine1982}. Quantum mechanics violates distance-based Bell inequalities and thus it does not obey this assumption. 

An example of an information-theoretic distance for binary observables $x_i=\pm 1$ is the covariance distance $d(X_i,X_j)=1-\langle X_i X_j\rangle$ introduced in \cite{Schumacher1991}. It uses correlations between $X_i$ and $X_j$. This distance is particularly interesting since it allows for a re-derivation of correlation-based Bell inequalities. Imagine that Alice and Bob share a bipartite system. Alice performs on her part one of two randomly chosen measurements, either $X_1$ or $X_3$, and Bob performs on his part either $X_2$ or $X_4$. Note, that due to the assumption we made before $ d \left( X_1, X_4  \right) \le  d \left( X_1, X_2  \right) + d \left( X_2, X_4  \right) $ and $ d \left( X_2, X_4  \right) \le  d \left( X_2, X_3  \right) + d \left( X_3, X_4  \right) $, even though $X_2$ and $X_4$ may not be jointly measurable. Combining these two triangle inequalities together results in a quadrangle inequality
\begin{align}\label{Nierownosc}
 d \left( X_1, X_4  \right) \le  d \left( X_1, X_2  \right) +  d \left( X_2, X_3  \right) +  d \left( X_3, X_4  \right).
\end{align}
Applying the covariance distance to the above inequality one obtains the well know  Clauser-Horne-Shimony-Holt (CHSH) inequality
\begin{align}
\langle X_1 X_2 \rangle+\langle X_2 X_3 \rangle+\langle X_3 X_4 \rangle - \langle X_1 X_4 \rangle \le 2. \nonumber
\end{align}

Here, we focus on information-theoretic distance measures based on entropies. In particular, we choose the following four examples
\begin{align}
\label{d1}d_{1} \left( X, Y  \right) &= H \left( X, Y  \right) - I \left( X, Y  \right), \\
\label{d2}d_{1'} \left( X, Y  \right) &= 1 - \frac{ I \left( X, Y  \right) }{ H \left( X, Y  \right)  } = \frac{ d_{1} \left( X, Y  \right)  }{ H \left( X, Y  \right) }, \\
\label{d3}d_{2}  \left( X, Y  \right) &= \max\{ H \left( X \right), H \left( Y \right) \} - I \left( X, Y  \right), \\
\label{d4}d_{2'}  \left( X, Y  \right) &= 1- \frac{ I \left( X, Y  \right) }{ \max\{ H \left( X \right), H \left( Y \right) \} } \\
&= \frac{ d_{2}  \left( X, Y  \right) }{ \max\{ H \left( X \right), H \left( Y \right) \} } \nonumber
\end{align}
already presented in \cite{Zurek1989} and \cite{Bennett1998}.  Here $H\left( X \right)$ is the entropy of $X$ and $I \left( X, Y  \right) =  H \left( X \right) + H \left( Y  \right) - H \left( X, Y  \right)$ is the mutual information between $X$ and $Y$. Note that $d_{1'(2')}$ is the normalized version of $d_{1(2)}$. 

Despite the fact that the above distances were defined for Shannon and Kolmogorov entropies, it was proven in \cite{Furuichi2006} that they are also valid information-theoretic distances if one uses Tsallis entropy for $q \geq 1$
\begin{align}\label{Tsallis}
H_{q}^{T} \left( X \right) = \frac{ 1 }{ q-1 } \left[ 1 -  \sum_{x} \left( p \left( X=x \right) \right)^q \right].
\end{align}
It should be noted that Tsallis entropy converges to Shannon entropy in the limit $q \rightarrow 1$
\begin{align}\label{Granica}
H^{Sh} \left( X \right) = - \sum_{x} p \left( X=x \right) \ln p \left( X=x \right)=\lim_{ q \rightarrow 1} H_{q}^{T}.
\end{align}

It may also seem natural to consider R\'enyi entropy in our discussion of Bell inequalities based on generalized entropies. However, unlike Tsallis entropy, the corresponding functions of R\'enyi entropy do not obey the triangle inequality.

\subsection{State}

We consider two qutrits in the following sate
\begin{align}
\rho_{\beta} = V | \psi \rangle \langle \psi |_{\beta} + \frac{1-V}{9} \mathbb{I}.
\end{align}
The parameter $V\in [0,1]$ is called the visibility, whereas $1-V$ is the amount of white noise added to the state
\begin{equation}\label{State}
| \psi \rangle_{\beta} = \frac{ 1 }{ \sqrt{2 + \beta^2 } }  \big( | 1,1  \rangle + | 2,2  \rangle + \beta | 3,3  \rangle \big),
\end{equation}
where $\beta \in [0,1]$. The reason to chose this state is the following. It is usually expected that a maximally entangled state should also maximally violate a tight Bell inequality; this is true for two entangled qubits but it does not hold for two entangled qutrtis in the state $|\psi\rangle_{\beta=1}$ as observed in \cite{Acin2005}. The optimal state happens to be $|\psi\rangle_{\beta\approx 0.792}$. This unusual behavior has not yet been satisfactorily explained. We would like to investigate if this discrepancy happens in entropic Bell inequalities.

\subsection{Measurements}

In this work we decide to represent measurements and the system using optical setups, which happens to give nice parametrization of the measurements. In this case Alice and Bob perform two randomly chosen local measurements each represented by a sequence of beam splitter and phase shifters as in \cite{Reck1994}, where the proposed experimental setups parametrizes unitary group of arbitrary $d$ dimensions. Each qudit from the entangled pair of two qudits encoded in the path of $d$ photons. In the general case the transformation done on each side can be represented as
\begin{align}
U(d) = \left( T_{d,d-1} \cdot T_{d,d-2} \cdot \ldots T_{2,1} \cdot D  \right)^{-1}.
\end{align}
Here matrix $T_{pq} = \left( t_{ij} \right)_{pq}$ denotes the operation of beamsplitter and phase shifters on $p$ and $q$ path modes. Its elements are $t_{pp} = e^{i \phi_{pq}} \sin \omega_{pq} $, $t_{qq} = - \sin \omega_{pq}$, $t_{pq} = e^{i \phi_{pq}}  \cos \omega_{pq}$, $t_{qp} = \cos \omega_{pq}$, the rest of diagonal terms equal to 1 and other off-diagonal elements are 0. Note that $T_{pq}$  can be realized by a standard Mach-Zehnder interferometer for two path modes. Subsequent transformations of $T_{pq}$ on all two-dimensional subspaces of the $d$-dimensional Hilbert space of the considered system together with additional phase shifts done by $D = \left( \delta_{ij} e^{i \alpha_i} \right) $ allows to encode any unitary operation \cite{Reck1994}.

The choice of all phase shifts determines the measurement settings of Alice (Bob)  $\{ \phi_{i_j}^{A (B)}, \omega_{i_j}^{A (B)}, \alpha_i^{A (B)}  \}$, $i=2,3$ and $j=1,2,\ldots,i$. 
The probability that Alice finds one photon in the $m$th mode on her side and Bob finds another in $n$th mode on his side
\begin{multline}
\mathrm{Prob} \left( m, n | \{ \phi_{i_j}^A, \omega_{i_j}^A, \alpha_i^A  \}, \{ \phi_{i_j}^B, \omega_{i_j}^B, \alpha_i^B  \} \right) \\ = \mathrm{Tr} \left( P_{mn}  \rho_{AB}' \right),
\end{multline}
where the state after unitary operation takes the form $\rho' = U_A \otimes U_B \rho U_A^{\dagger} \otimes U_B^{\dagger}$ and $P_{mn}  =| m, n \rangle \langle m, n |$ describes the projective measurement on each side.

Note, that local probabilities of Alice (Bob) do not depend on Bob's (Alice's) settings and can be obtained as marginals of $\mathrm{Prob} \left( m, n | \{ \phi_{i_j}^A, \omega_{i_j}^A, \alpha_i^A  \}, \{ \phi_{i_j}^B, \omega_{i_j}^B, \alpha_i^B  \} \right)$
\begin{multline}
\mathrm{Prob} \left( m | \{ \phi_{i_j}^A, \omega_{i_j}^A, \alpha_i^A  \} \right) \\= \sum_n \mathrm{Prob} \left( m, n | \{ \phi_{i_j}^A, \omega_{i_j}^A, \alpha_i^A  \}, \{ \phi_{i_j}^B, \omega_{i_j}^B, \alpha_i^B  \} \right). 
\end{multline}

\begin{figure}[t] 
\centering 
\includegraphics[scale=0.45]{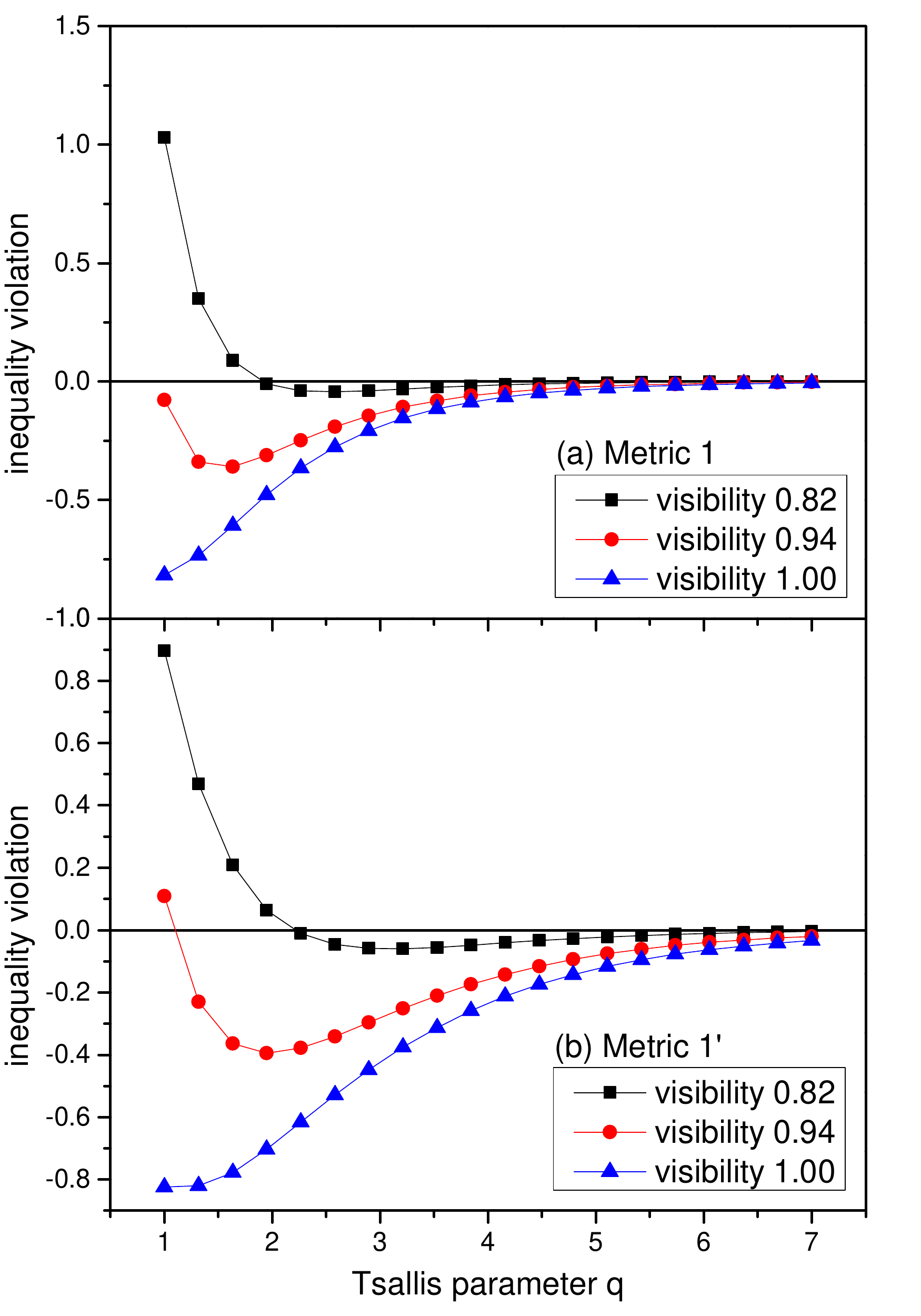} 
\caption{\label{fig:Rys1} Violations of inequalities constructed upon metrics (\ref{d1}) and (\ref{d2}), $\beta=0$.}
\end{figure}

\begin{figure}[t] 
\centering 
\includegraphics[scale=0.45]{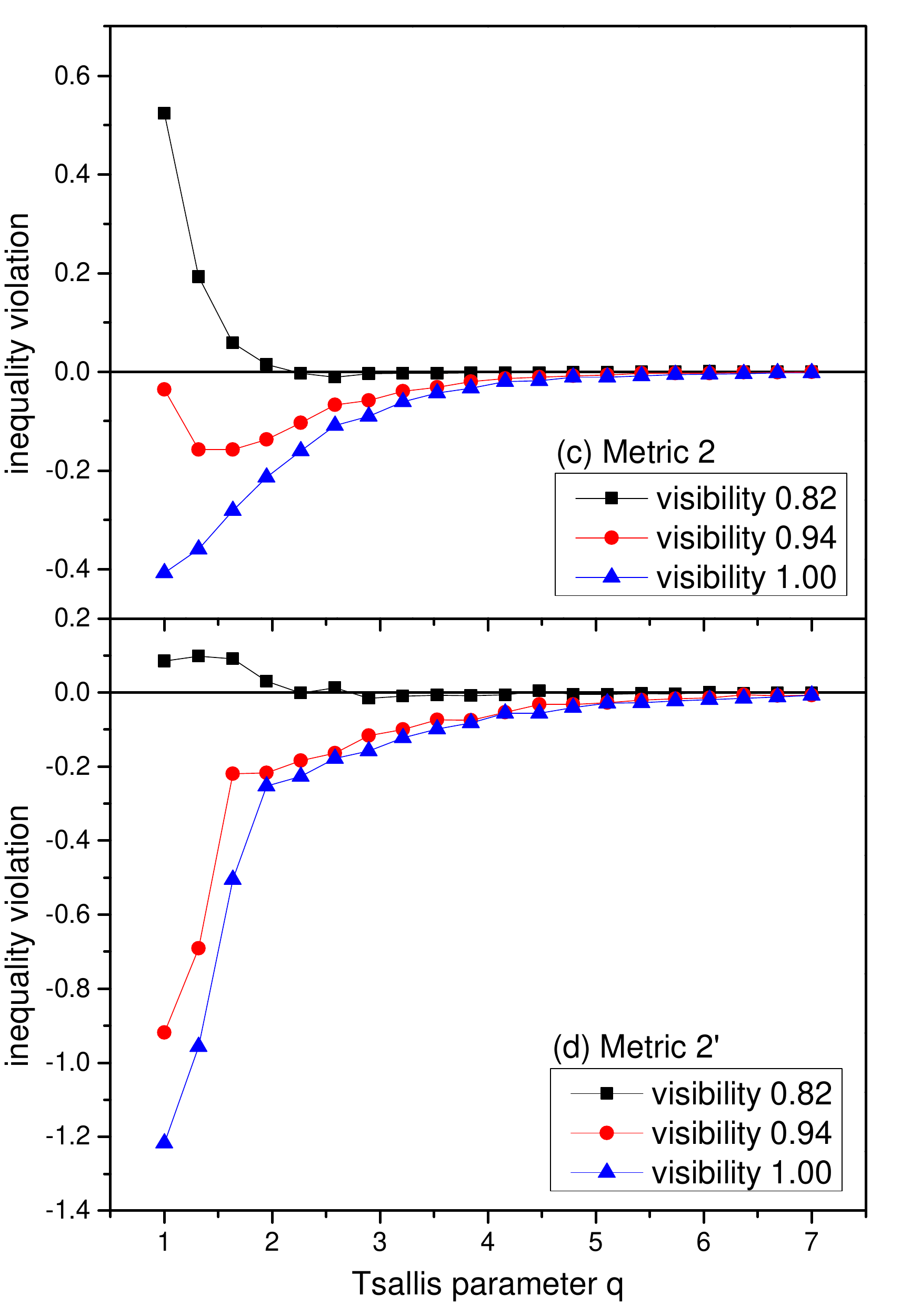} 
\caption{\label{fig:Rys2} Similar to Fig. (\ref{fig:Rys1}). In this case the inequalities are constructed upon metrics (\ref{d3}) and (\ref{d4}), $\beta=0$.}
\end{figure}

\subsection{The inequality}

The inequality considered in this work is based on the quadrangle inequality (\ref{Nierownosc}) 
\begin{align}\label{Nierownosc2}
 d \left( A, B'  \right) \le  d \left( A, B  \right) +  d \left( B, A'  \right) +  d \left( A', B'  \right),
\end{align}
where A, A', B and B' are different settings of Alice and Bob that are determined by the phase shifts $\{ \phi_{i_j}^A, \omega_{i_j}^A, \alpha_i^A  \}$, $\{ \phi_{i_j}^{A'}, \omega_{i_j}^{A'}, \alpha_i^{A'}  \}$, $\{ \phi_{i_j}^B, \omega_{i_j}^B, \alpha_i^B  \}$ and $\{ \phi_{i_j}^{B'}, \omega_{i_j}^{B'}, \alpha_i^{B'}  \}$, respectively. We investigate this inequality for the four metrics (\ref{d1}-\ref{d4}) in which all entropies and mutual information are taken with respect to Eq. (\ref{Tsallis}), i.e., information-theoretic metrics are based on Tsallis entropies that are functions of $\mathrm{Prob} \left( m, n | \{ \phi_{i_j}^A, \omega_{i_j}^A, \alpha_i^A  \}, \{ \phi_{i_j}^B, \omega_{i_j}^B, \alpha_i^B  \} \right)$, $\mathrm{Prob} \left( m | \{ \phi_{i_j}^A, \omega_{i_j}^A, \alpha_i^A  \} \right)$ and $\mathrm{Prob} \left( n | \{ \phi_{i_j}^B, \omega_{i_j}^B, \alpha_i^B  \} \right)$.

\section{Numerical results}

\begin{figure}
\centering
\includegraphics[scale=0.35]{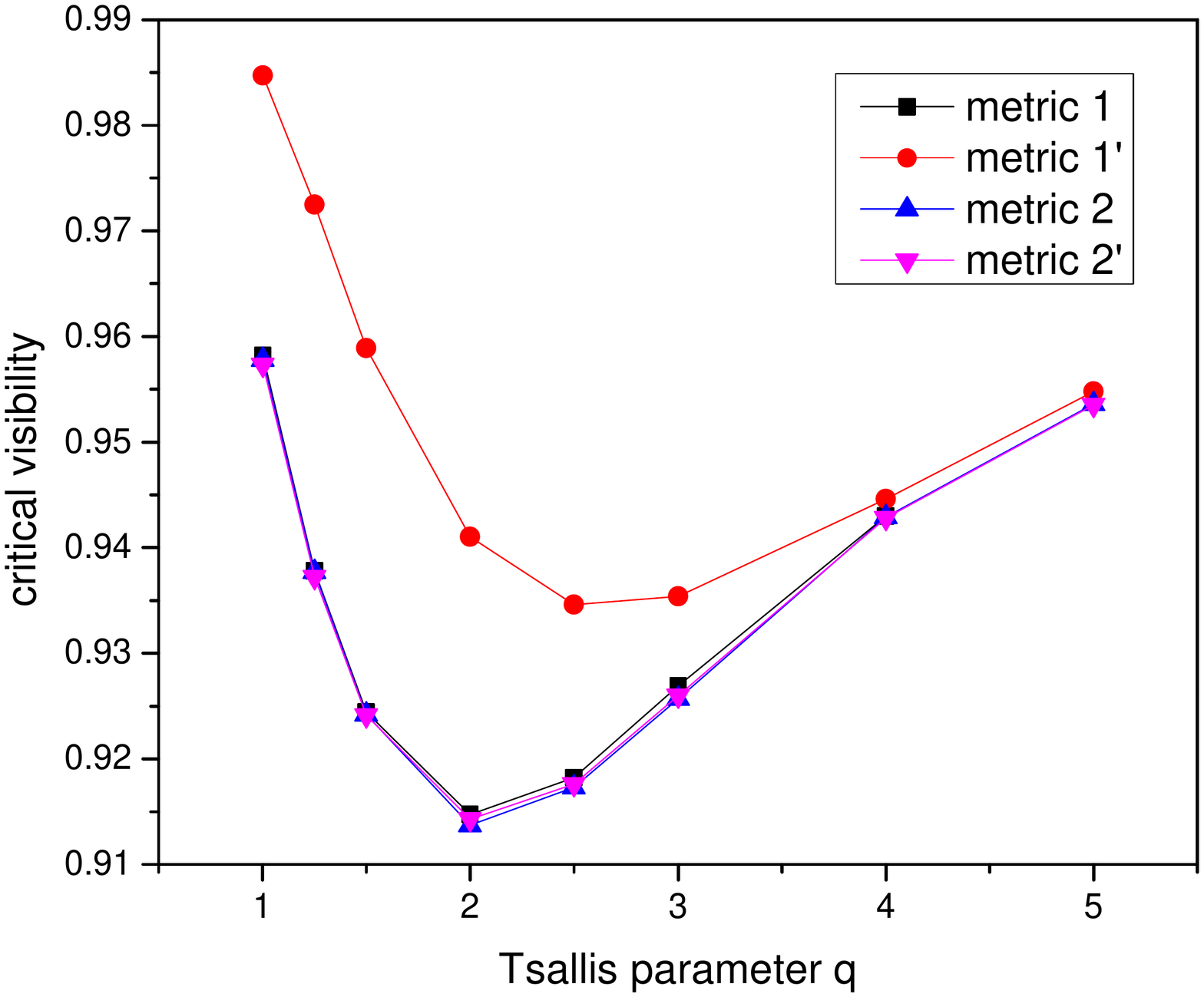} 
\caption{\label{fig:Rys3} The dependance of visibility $V_c$ on Tsallis parameter $q$ for $\beta=1$. The relation is the same for metric (\ref{d1}), (\ref{d3}), and (\ref{d4}).}
\includegraphics[scale=0.35]{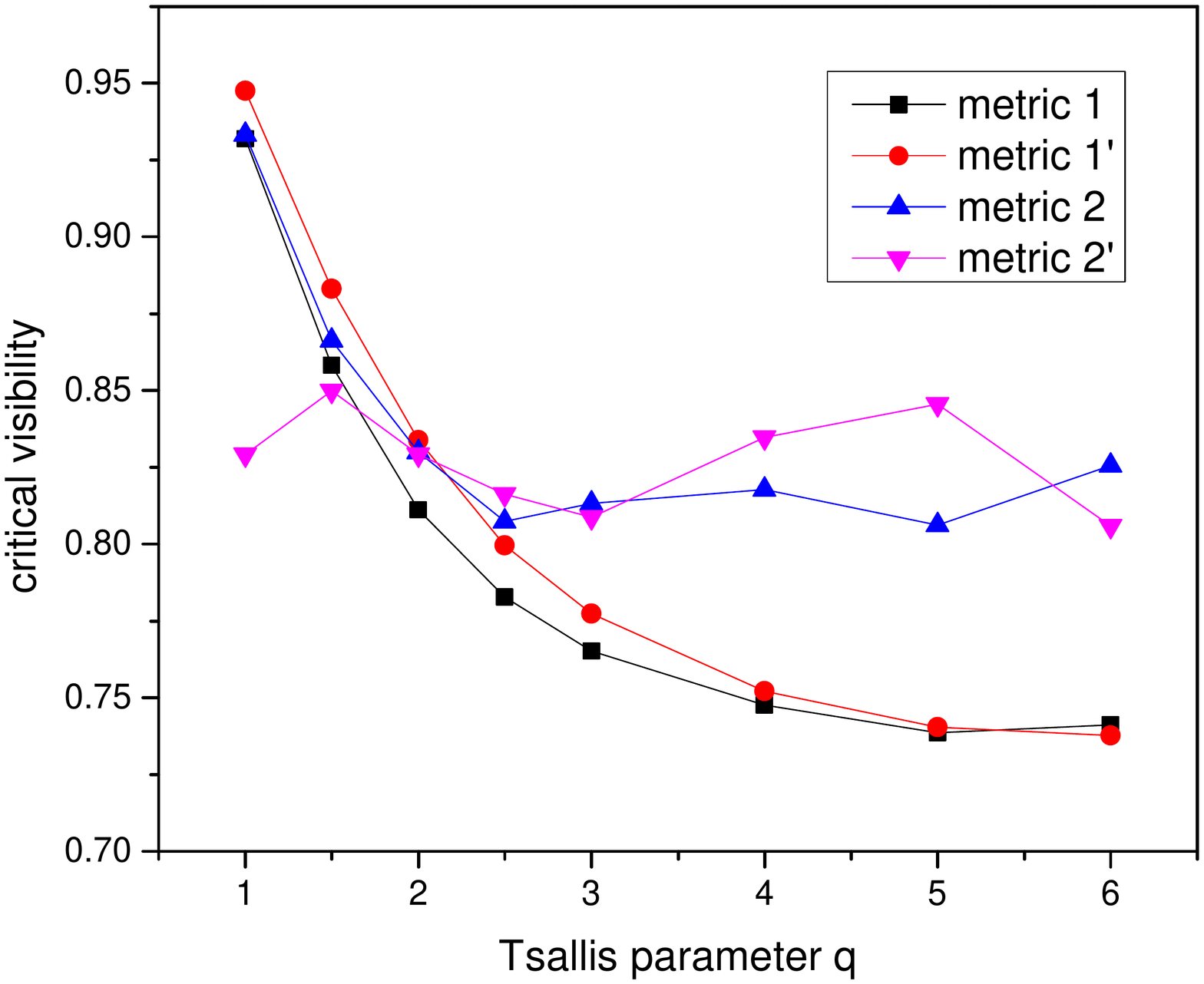} 
\caption{\label{fig:Rys4} The dependance of visibility $V_c$ on Tsallis parameter $q$ for $\beta=0$.}
\end{figure}

We consider $\rho$ based on entangled qutrits with additional parameter $\beta$. The plotted data was obtained for metrics (\ref{d1})-(\ref{d4}) constructed upon Tsallis entropy. Violation of (\ref{Nierownosc2}) occurs when left-hand side ($L$) of (\ref{Nierownosc2}) is greater than right-hand side ($R$). Therefore we call the quantity $R-L$ the inequality violation. Presented data in Fig. (\ref{fig:Rys1})-(\ref{fig:Rys3})  was generated in the program finding minimum of $R-L$ over all phases $\{ \phi_{i_j}^A, \omega_{i_j}^A, \alpha_i^A  \}$, $\{ \phi_{i_j}^B, \omega_{i_j}^B, \alpha_i^B  \} \in [0, 2\pi]$ for Alice and Bob with fixed $\beta$. In first runs the optimization procedure was done with free $\beta$ but each time we were finding the minimum for $\beta$ close to zero and therefore we decided to keep it fixed in order to make this numerical problem less complex.

From the relation (\ref{Granica}) it follows that in Fig. (\ref{fig:Rys1}) and Fig. (\ref{fig:Rys2}) for Tsallis parameter $q \rightarrow 1$, i. e. the very first points for each series, we find the value of inequality violation for the distance measure based on Shannon entropy. We observed that for sufficiently high noise parameter Tsallis entropy for $q>$1 gives better violation than Shannon entropy (see Figs. (\ref{fig:Rys1}) and (\ref{fig:Rys2})). It is also very important to point out the presence of violation detected by Tsallis entropy whereas the Shannon's does not indicate any non-classical behavior. In fact, for example for metric $d_{1'}$ (\ref{d2}) we see that there is an indication of violation for the state with visibility $V=0.94$ but only in the case of Tsallis entropy.

For better check the dependance on noise for maximally entangled state, we looked at the behavior of critical visibility $V_c$. It is defined as the smallest value of $V$ for which we still observe inequality violation. In the best case for shannon entropy  the lowest value is $V_c=0.96$ whereas with Tsallis' we can go up to $0.915$ for $\beta=1$ (Fig. \ref{fig:Rys3}). 

As mentioned before, if we release the constraint $\beta=1$ we observe that the strongest violations occurs for $\beta$ close to zero. Indeed, results that we get for $\beta = 0$ show very interesting behavior of critical visibility which in that case goes down to $0.71$  (Fig. \ref{fig:Rys4}), i.e. much lower that in the previous instances.

The key point to understand why Tsallis entropy is more suitable lies in the parameter $q$. The definition (\ref{Tsallis}) can be interpreted as the formula of probabilities with some weights governed by $q$. By tuning it we can make events of lower probability much less important and focus only on most likely instances \cite{Maszczyk2008}.

\section{Conclusions}

For the states with some amount of noise we have suggested to use Tsallis entropy rather than Shannon's as it allows to find stronger violations of entropic Bell-type inequalities. In our numerical simulations we found that $\beta $ is close to zero which means that for the considered inequality derived from triangle principle the maximal violations occurs for the non-maximally entangled state.

\section{Acknowledgements}

This work is supported by the National Research Foundation and Ministry of Education in Singapore. In addition, P. K. and D. K. are supported by the Foundational Questions Institute (FQXi).

\bibliography{Artykul.bib}

\end{document}